# Broadband highly efficient nonlinear optical processes in on-chip integrated lithium niobate microdisk resonators of Q-factor above 10^8

Renhong Gao[1,6,10], Haisu Zhang[2,3,10], Fang Bo[4], Wei Fang[5], Zhenzhong Hao[4], Ni Yao[5], Jintian Lin[1,†], Jianglin Guan[2,3], Li Deng[2,3], Min Wang[2,3], Lingling Qiao[1], and Ya Cheng[1,2,3,6,7,8,9,*]

[1]*State Key Laboratory of High Field Laser Physics and CAS Center for Excellence in Ultra-intense Laser Science, Shanghai Institute of Optics and Fine Mechanics (SIOM), Chinese Academy of Sciences (CAS), Shanghai 201800, China.*
[2]*XXL—The Extreme Optoelectromechanics Laboratory, School of Physics and Electronic Science, East China Normal University, Shanghai 200241, China.*
[3]*State Key Laboratory of Precision Spectroscopy, East China Normal University, Shanghai 200062, China.*
[4]*The MOE Key Laboratory of Weak Light Nonlinear Photonics, TEDA Applied Physics Institute and School of Physics, Nankai University, Tianjin 300457, China.*
[5]*State Key Laboratory of Modern Optical Instrumentation, College of Optical Science and Engineering, Zhejiang University, Hangzhou 310027, China.*
[6]*University of Chinese Academy of Sciences, Beijing 100049, China.*
[7]*Collaborative Innovation Center of Extreme Optics, Shanxi University, Taiyuan 030006, China.*
[8]*Collaborative Innovation Center of Light Manipulations and Applications, Shandong Normal University, Jinan 250358, China.*
[9]*Shanghai Research Center for Quantum Sciences, Shanghai 201315, China.*
[10]*These authors contributed equally to the work.*

[†]*jintianlin@siom.ac.cn*

[*]*ya.cheng@siom.ac.cn*

(Dated: March 11, 2021)

# Abstract

We demonstrated broadband highly efficient optical nonlinear processes in on-chip integrated lithium niobate (LN) microdisk resonators. The Q factors of the micro-resonators fabricated by femtosecond laser writing and chemo-mechanical polishing are reliably above $10^8$, approaching the intrinsic material absorption limit of LN. Broadband nonlinear processes, including optical parametric oscillation (OPO), second harmonic generation (SHG), third harmonic generation, and fourth harmonic generation, were observed with ultrahigh efficiencies in the same LN microdisk without introducing domain inversion, thanks to the natural quasi phase-matching and the dense spectral modes of the X-cut LN microdisk with millimeter diameter. The threshold of OPO and the absolute conversion efficiency of SHG are 19.6 μW and 66%, both surpass the state-of-the-art values among on-chip LN micro-resonators demonstrated so far. The broadband and highly efficient nonlinear frequency conversions achieved with the ultrahigh-Q LN microdisk resonators promise high-density integration of nonlinear photonic devices such as frequency convertors and entangled photon sources.

Whispering gallery mode (WGM) optical resonators are highly desirable in a broad range of fields including nonlinear optics [1,2], cavity optomechanics [3,4], quantum information [5,6] and lightwave communications [7]. Light loaded into WGM-resonators is sharply confined by the continuous total internal reflection around the interface of the volume with its surroundings, facilitating the long-time storage of optical power at specific resonant frequencies as well as the large enhancement of nonlinear optical interactions, both demanding higher Q-factors of the resonators [8]. Due to rapid progress of photonic integrated circuits (PICs) thin-film wafer based WGM micro-resonators have boomed, among which the high-Q ($10^5 \sim 10^7$) micro-resonators fabricated on thin film lithium niobate (LN) have attracted broad interests due to the excellent electro-optic and nonlinear optical properties of the underlying materials [9,10]. Intriguing applications of frequency comb generation [11-13], microwave-to-optical transducer [14], quantum light source [5,15], and single photon nonlinearity [16] have been demonstrated in high-Q LN WGM-resonators.

On the other hand, besides the ultrahigh Q factors, phase-matchings between the pertinent waves underpin efficient and low-threshold nonlinear frequency conversions in WGM-resonators [2]. Modal phase-matching (MPM) by utilizing geometric dispersions and quasi phase-matching (QPM) by patterning the nonlinearity of materials are normally employed for frequency conversions in on-chip micro-resonators [17-22]. MPM necessitates the use of both fundamental and high-order WGM modes where a reduced spatial overlap between these modes lessens the effective nonlinearities. QPM profits from all involved waves residing in the fundamental modes, though extra complexity and loss during the nonlinearity patterning process compromise the attainable Q-factors [22]. Besides, the nonlinearity patterning is conjugated with the specific nonlinear process whose compatibility for other/cascaded nonlinear processes have to be delicately controlled by exerting complicated techniques such as dual/multi/chirped-periods spatial patterning for flexible QPMs [23-25], though the fabrication processes inevitably induce some fluctuations rendering the produced micro-resonators deviating from the perfectly resonant conditions for frequency (energy) and phase (momentum) conservations. An additional tuning method (normally by thermal-optic effect) is thus required to restore the micro-resonators to the desired resonant conditions [19-23]. However, this thermal recovery is rather frustrated in ultrahigh-Q (UHQ, >$10^8$) on-chip micro-resonators due to the ultra-narrow linewidth of the resonant modes. As a result, both MPM and QPM are usually employed in micro-resonators with mediately high Q-factors ($10^5$~$10^7$) for specific frequency conversion with limited flexibility and bandwidth.

Recently, a distinct phase-matching scheme dubbed as natural quasi-phase matching (NQPM) has been widely investigated [26]. The inherent cyclic phase-matching for transverse-electric (TE) WGMs circulating along the circumstance of the $\chi^{(2)}$ micro-resonator with in-plane optical axis naturally facilitates broadband and efficient frequency conversions without resting on specific modal dispersion and nonlinearity pattering [26,27]. Specifically, the "natural" phase-matching character independent of additional tuning and patterning makes NQPM promising in UHQ (>$10^8$) $\chi^{(2)}$ micro-resonators for broadband and versatile phase-matched frequency conversions, paving the road for reconfigurable frequency convertors and quantum light sources. However, such kind of investigation is still missing so far, and diverse nonlinear processes except second harmonic generation (SHG) have not yet been demonstrated through NQPM in $\chi^{(2)}$ micro-resonators.

In this work, by simultaneous utilizations of an optimized micro-fabrication method for producing UHQ (>$10^8$) microdisk resonators in the pristine LN thin film and the NPQM without additional patterning of nonlinearity (periodic poling and domain inversion for LN) in the fabricated microdisk resonators, broadband frequency conversions including SHG, optical parametric oscillation (OPO), third harmonic generation (THG), and fourth harmonic generation (FHG), were observed with ultrahigh efficiencies at the same LN micro-resonator. The threshold of OPO as low as 19.6 µW and the absolute conversion efficiency of SHG at 0.11 mW pump power as high as 66%, have been demonstrated, outperforming the best reported values in on-chip LN micro-resonators [19,20,28]. This record-low threshold (~20 µW) for the OPO process, holds great promise for on-chip integrated bright photon source by spontaneous parametric down-conversion in the LN microdisk resonators.

The LN microdisk resonators were fabricated on the LN thin film wafer prepared by crystal cutting, polishing and bonding. Femtosecond laser direct writing and chemo-mechanical polishing was then adopted for shaping the LN thin film into free-standing microdisk resonators (fabrication details can be found in the Supplemental materials). The main advantage of our fabrication method relies on the optimized use of chemo-mechanical polishing instead of usually adopted ion-implantation and plasma etching for LN thin film preparation and etching, essentially alleviating the crystalline damages and surface/interface defects during these processes and achieving a superior interior and interface quality of the LN microdisk resonator having the Q-factors approaching the intrinsic material absorption limit (giving Q~$2\times10^8$ at 1550 nm wavelength [29]).

The scanning-electron-microscopy (SEM) image of the fabricated LN microdisk resonator with a diameter of 1030 µm was shown in Figs. 1(a-b). The average surface roughness of the resonator was less than 0.5 nm [30,31]. The wedge angle of the microdisk sidewall is 8°, rendering the thickness of the microdisk resonator being varied from the submicron-scale at the edge to 3.5 µm near the center, as shown in Fig. 1(b). The fundamental WGM is simulated and plotted in the top inset of Fig. 1(b), with the center of the mode located 12 µm horizontally from the microdisk edge where the local LN thin film is only 600 nm thick. The distance of the supporting $SiO_2$ pillar to the microdisk edge was chosen to be 100 µm (see bottom inset of Fig. 1(b)) by controlling the HF acid etching time, which is sufficient to protect the WGMs from the scattering loss induced by the rough $SiO_2$ pillar. As a consequence, the high-quality LN thin film with ultra-smooth interface/surface, the clear isolation of the optical modes from the undercut $SiO_2$ together with the tightly guided modes in submicron thick LN thin films, all enabled by the optimized fabrication process, bring about UHQ factors as well as small mode volumes of the fabricated micro-resonators ideal for efficient frequency conversions as shown below.

The Q-factor of the LN resonator was first measured by scanning the wavelength of a narrow-linewidth laser (<200 kHz) injected into the resonator with the help of a fiber taper. The relative position between the fiber taper and the resonator can be finely adjusted to obtain critical coupling. Low input power of 5 µW was used to avoid the thermo-optic effect and nonlinear optical processes. From the measured transmission spectrum plotted in Fig. 1(c), a typical Lorentzian-shaped resonance profile centered at ~1551.52 nm is fitted and shown in the inset of Fig. 1(d), indicating a loaded (intrinsic) Q factor of $7.5 \times 10^7$ ($1.20 \times 10^8$). Cavity ring-down measurement was also implemented for precise characterization of the Q-factor by repeatedly scanning the laser with a high-speed Mach-Zehnder modulator into the targeted resonance, and the result was shown in Fig. 1(d). The retrieved photon lifetime is 64.3 ns, giving the intrinsic Q factor as high as $1.23 \times 10^8$, which agrees well with the transmission spectrum measurement. This intrinsic Q factor for the 1030 µm diameter LN microdisk resonator is the highest among the demonstrated on-chip LN micro-resonators so far [9], and smaller microdisks with diameters of 130 µm and 240 µm were also fabricated showing intrinsic Q factors both of $0.9 \times 10^8$. (The details of the results are provided in the Supplemental materials). Since the bending losses are exponentially inverse proportional to diameters of the resonators, the tiny variance (<30%) of the Q-factors for microdisk resonators with diameters from ~1 mm to ~100 µm indicates the obtained Q-factors are solely determined by the intrinsic material absorption.

Frequency conversions in the UHQ LN microdisk resonators were further investigated. To maintain the UHQ factor without introducing the domain inversion like QPM, X-cut LN microdisk resonators were employed for achieving the NQPM as revealed recently [26]. The optical axis is aligned within the plane of the X-cut LN microdisk, rendering both the effective refractive indexes and nonlinear coefficients of the excited TE WGMs oscillate when the guided light is circulating along the circumstance of the X-cut LN microdisk. By controlling the polarization of the injected pump light to excite the TE mode of the X-cut LN resonator, strong SHG and THG signals were observed when the wavelength of the pump light was set as 1555.2 nm. The spectrum was recorded by an optical spectrum analyzer (OSA, detected wavelength range of 600~1700 nm) and a fiber spectrometer (detected wavelength range of 300~1000 nm) and shown in Fig. 2(a). The second harmonic and third harmonic signals were both TE polarized as well. The output powers of the harmonics were collected by an objective lens with a numerical aperture of 0.42 and sent into a power meter. The output power of SHG grows linearly with the square of the in-coupled pump power with a very high normalized conversion efficiency of 602%/mW, as shown in Fig. 2(b). The absolute conversion efficiency was as high as 66% when the in-coupled pump power was raised to 0.11 mW, and this value is also much greater than the 15% conversion efficiency of the periodically poled LN (PPLN) microring at similar pump powers [19]. Besides, the normalized conversion efficiency of THG was measured to be 30%/mW$^2$, outperforming the best reported values as well [21,26,32], as shown in Fig. 2(c).

All these ultrahigh conversion efficiencies benefit from the realizations of UHQ factor, small mode volume V, high nonlinear optical coefficient ($d_{33}$ for LN) and NQPM simultaneously achieved with the on-chip X-cut LN microdisk resonators. Meanwhile, due to the dense spectral mode distribution of the millimeter-diameter resonator, broadband NQPM can be anticipated for a plethora of WGM pairs. To check this hypothesis, the transient output powers of harmonics were detected by continuously tuning the pump wavelength from 1530 nm to 1570 nm with a speed of 0.5 nm/s, and the corresponding output powers of SHG and THG at each pump wavelength are shown in Fig. 2(d). The resolution of the response curves was limited by the power meter with a sampling rate of 10 Hz. Remarkably, a broadband NQPM for both SHG and cascaded THG, covering the full telecom C-band with conversion efficiencies close to that at 1555.2 nm pump wavelength (~600%/mW), was clearly observed. Although the normalized SHG conversion efficiency of our ultrahigh Q LN microresonators is lower than that of the PPLN microring achieved recently [16], the concurrent broadband SHG response in our microresonators prevails over PPLN microrings in situations demanding both high conversion efficiency and wide bandwidth.

When the in-coupled pump power raised to more than 0.11 mW, higher-order nonlinearities such as Raman scattering and Kerr effect emerged. These processes impeded the increase of harmonic, and induced the oscillation of the output power of harmonic other than drove SHG into saturation state. Figure 3(a) shows clearly visible light emission from the resonator, which was captured by a visible CCD camera. The spectrum ranging from 385 to 1700 nm wavelength recorded by the OSA (red line) and the fiber spectrometer (black line) is plotted in Fig. 3(b), when the in-coupled power was 3.8 mW. Here, FHG from the microdisk without sophisticated dual-periodically polling [21] was also detected at 388.8 nm wavelength with an output power as high as 3.16 μm. The Raman assisted frequency comb generation was observed around pump wavelength with a comb line spacing of 7.9 THz, as shown in the rectangular box II in Fig. 3(b). Cascaded Raman assisted frequency comb generation was also detected around second harmonic wavelength for the first time, which is shown in the rectangular box I in Fig. 3(b). Each comb line in box I was generated by sum frequency generation process between the fundamental light and the comb line in box II. There are also a spectrum of cascaded Raman scattering lines [31,33-35] appearing from 388 nm to more than 1700 nm (limited by the detected range of the OSA). The large spectral emission is demonstrated with only one pump light, which is due to the combination of UHQ factor and the excellent nonlinear optical properties of LN.

When the pump laser was replaced with a near-visible tunable laser (765-781 nm) above certain threshold power, broadband non-degenerated OPO signals would be observed. Figure 4(a) shows such an OPO spectrum detected by the OSA when pumping at 770.4 nm wavelength with 20.6 μW in-coupled power. Signal wave and idler wave with wavelengths of 1517.2 nm and 1565.2 nm, were generated from the pump wave, satisfying the phase matching and triply resonant condition. The detected output power of OPO increases linearly as the increasing pump power above certain threshold power, as shown in Fig. 4(b). The threshold power was determined as low as 19.6 μW, which is lower than the best previously reported value (93 μW) in on-chip LN micro-resonator [28]. The gain rate of OPO was fitted as 20% by linear fitting. Such a low threshold was resulted from the combination of UHQ factor and the using of NQPM scheme. Moreover, the NQPM scheme allows broadband phase matching, which has also been experimentally confirmed. For example, when the pump wavelength was tuned to be 776.5 nm, OPO waves were detected with the threshold power of 22.5 μW, as shown in Fig. 4(c).

The versatility of broadband NQPM for a wide range of nonlinear optical processes, including SHG, THG, FHG, and OPO, realized on the same microdisk resonator is quite appealing for further investigations. Previous works have unveiled the indispensable role of the natural "poling" of the effective nonlinearity in X-cut LN microdisk on mitigating the wave-vector mismatch ($\Delta k$) which is also oscillating due to the anisotropic refractive indices along the periphery of the microdisk [26,27]. Specifically, the oscillation of $\Delta k$ relaxes the stringent requirement by QPM at unique poling period. Besides, NQPM inherently contains four periods of nonlinearity for QPM, and the periods are determined by the diameter of the microdisk which can be easily controlled (see Fig. 4(d) as well as the detailed derivations in the Supplemental materials). Moreover, since the spectral mode density of the microdisk depends on its diameter as well, the millimeter-scale microdisk employed in this work can provide plenty sets of WGMs with suitable wave-vector mismatches covered by the small momentum (long period) of the nonlinearity poling, as corroborated by the simultaneous observations of efficient and broadband frequency conversions arising from various nonlinear processes with distinct phase-matching criteria. It should be noted that although NQPM entails the use of high-order WGMs which reduces the effective nonlinearity, its advantage in the UHQ micro-resonators can easily compensate for this defect by significantly enhanced resonance and achieve ultrahigh efficiencies for multiple nonlinear optical processes.

To summarize, the on-chip UHQ factor in excess of $10^8$ on pristine LN thin film wafer is experimentally demonstrated for the first time. Broadband SHG, THG, and FHG with ultrahigh normalized conversion efficiencies and low-threshold OPO were observed at the same LN microdisk resonator without introducing domain inversion. The combination of UHQ photonic devices and excellent nonlinear optical property, together with high-speed electro-optic modulation of LN, will play significant roles in integrated quantum information processing, modern commutations, and cavity quantum electrodynamics.

The work is supported by the Ministry of Science and Technology of China (Grant No.2019YFA0705000), the NSFC (Grants No.11734009, 11874375, 11874154), Key Research Program of Frontier Sciences (QYZDJ-SSWSLH010), Strategic Priority Research Program of Chinese Academy of Sciences (XDB16030300), Shanghai Municipal Science and Technology Major Project (2019SHZDZX01), Higher

Education Discipline Innovation Project (B07013), and the Youth Innovation Promotion Association of Chinese Academy of Sciences (Grant No. 2020249).

**References:**

[1] X. Zhang, Q.-T. Cao, Z. Wang, Y.-x. Liu, C.-W. Qiu, L. Yang, Q. Gong, and Y.-F. Xiao, Nat. Photon. **13**, 21 (2019).

[2] G. Lin G, A. Coillet, and Y. K. Chembo, Adv. Opt. Photon. **9**, 828 (2017).

[3] Z. Shen, Y.-L. Zhang, Y. Chen, C.-L. Zou, Y.-F. Xiao, X.-B. Zou, F.-W. Sun, G.-C. Guo, and C.-H. Dong, Nat. Photon. **10**, 657 (2016).

[4] T. J. Kippenberg and K. J. Vahala, Science **321**, 1172 (2008).

[5] Z. Ma, J.-Y. Chen, Z. Li, C. Tang, Y. M. Sua, H. Fan, and Y.-P. Huang, Phys. Rev. Lett. **125**, 263602 (2020).

[6] A. Fülöp, M. Mazur, A. Lorences-Riesgo, Ó. B. Helgason, P.-H. Wang, Y. Xuan, D. E. Leaird, M. Qi, P. A. Andrekson, A. M. Weiner, and V. Torres-Company, Nat. Commun. **9**, 1598 (2018).

[7] A. Guarino, G. Poberaj, D. Rezzonico, R. Degl'Innocenti, and P. Günter, Nat. Photon. **1**, 407 (2007).

[8] K. J. Vahala, Nature **424**, 839 (2003).

[9] J. Lin, F. Bo, Y. Cheng, and J. Xu, Photon. Res. **8**, 1910 (2020).

[10] Y. Jia, L. Wang, and F. Chen, Appl. Phys. Rev. **8**, 011307 (2021).

[11] Y. He, Q.-F. Yang, J. Ling, R. Luo, H. Liang, M. Li, B. Shen, H. Wang, K. Vahala,

and Q. Lin, Optica **6**, 1138 (2019).

[12] Z. Gong Z, X. Liu, Y. Xu, and H. X. Tang, Optica **7**, 1275 (2020).

[13] C. Wang, M. Zhang, M. Yu, R. Zhu, H. Hu, and M. Lončar, Nat. Commun. **10**, 1 (2019).

[14] M. Soltani, M. Zhang, C. Ryan, G. J. Ribeill, C. Wang, and M. Loncar, Phys. Rev. A **96**, 043808 (2017).

[15] R. Luo, H. Jiang, S. Rogers, H. Liang, Y. He, and Q. Lin, Opt. Express **25**, 24531 (2017).

[16] J. Lu, M. Li, C.-L. Zou, A. A. Sayem, and H. X. Tang, Optica **7**, 1654 (2020).

[17] R. Luo, Y. He, H. Liang, M. Li, J. Ling, and Q. Lin, Phys. Rev. Appl. **11**, 034026 (2019).

[18] X. Ye, S. Liu, Y. Chen, Y. Zheng, and X. Chen, Opt. Lett. **45**, 523 (2020).

[19] J. Lu, J. B. S., X. Liu, A. W. Bruch, Z. Gong, Y. Xu, and H. X. Tang, Optica **6**, 1455 (2019).

[20] J.-Y. Chen, Z.-H. Ma, Y. M. Sua, Z. Li, C. Tang, and Y.-P. Huang, Optica **6**, 1244 (2019).

[21] L. Zhang, Z. Hao, Q. Luo, A. Gao, R. Zhang, C. Yang, F. Gao, F. Bo, G. Zhang, and J. Xu, Opt. Lett. **45**, 3353 (2020).

[22] Z. Hao, L. Zhang, A. Gao, W. Mao, X. Lyu, X. Gao, F. Bo, F. Gao, G. Zhang, and J. Xu, Sci. China-Phys. Mech. Astron. **61**, 114211 (2018)

[23] S. Zhu, Y. Zhu, and N. Ming, Science **278**, 843 (1997).

[24] T. Beckmann, H. Linnenbank, H. Steigerwald, B. Sturman, D. Haertle, K. Buse, and I. Breunig, Phys. Rev. Lett. **106**, 143903 (2011).

[25] D. Haertle, J. Opt. **12**, 035202 (2010).

[26] J. Lin, N. Yao, Z. Hao, J. Zhang, W. Mao, M. Wang, W. Chu, R. Wu, Z. Fang, L. Qiao, W. Fang, F. Bo, and Y. Cheng, Phys. Rev. Lett. **122**, 173903 (2019).

[27] A. Lorenzo-Ruiz and Y. Léger, ACS Photonics **7**, 1617 (2020).

[28] J. Lu, A. A. Sayem, Z. Gong, J. B. Surya, C.-L. Zou, H. X. Tang, arXiv:2101.04735 (2021).

[29] V. S. Ilchenko, A. A. Savchenkov, A. B. Matsko, and L. Maleki, Phys. Rev. Lett. **92**, 043903 (2004).

[30] J. Zhang, Z. Fang, J. Lin, J. Zhou, M. Wang, R. Wu, R. Gao, and Y. Cheng, Nanomaterials **9**, 1218 (2019).

[31] R. Wu, J. Zhang, N. Yao, W. Fang, L. Qiao, Z. Chai, J. Lin, and Y. Cheng, Opt. Lett. **43**, 4116 (2018).

[32] K. Sasagawa and M. Tsuchiya, Appl. Phys. Express **2**, 122401 (2009).

[33] R. Wolf, I. Breunig, H. Zappe, and K. Buse, Opt. Express **25**, 29927 (2017).

[34] Z. Fang, H. Luo, J. Lin, M. Wang, J. Zhang, R. Wu, J. Zhou, W. Chu, T. Lu, and Y. Cheng, Opt. Lett. **44**, 5953 (2019).

[35] M. Yu, Y. Okawachi, R. Cheng, C. Wang, M. Zhang, A. L. Gaeta, and M. Lončar, Light Sci. Appl. **9**, e9 (2020).

**Captions of figures:**

Fig. 1 (Color online) The microdisk and Q factor measurement. (a) SEM of the microdisk. (b) The zoom-in SEM of the sidewall from the side view, upper inset: the electric field distribution, and lower inset: The optical micrograph of the resonator, showing a relatively small pillar whose rough boundary is far from the periphery of the resonator with more than 100 μm, where the scale bar is 200 μm. (c) The transmission spectrum from the fiber taper coupled with the resonator. (d) Ring-down measurement of the mode at 1551.52 nm wavelength, showing a lifetime τ of 64.3 ns, Inset: The resonance linewidth as narrow as 0.0207 pm, corresponding to loaded Q of $7.5 \times 10^7$.

Fig. 2 (Color online) Harmonic generation from the resonator via NQPM. (a) The spectra of pump light, SHG and THG. Power dependence of (b) SHG and (c) THG. (d) Pump-wavelength dependence of the output powers of SHG and THG.

Fig. 3 (Color online) (a) Clearly visible optical emission with near infrared pump of 1555.2 nm wavelength. Inset: Optical micrograph of the visible emission from the edge of the resonator from the sideview. (b) Spectrum of the emission.

Fig. 4 (Color online) (a) The spectrum of OPO, Inset: Optical micrograph of the microdisk pumped with 770.4 nm wavelength laser. (b) Power dependence of OPO, showing a threshold pump power of ~19.6 μW and a gain rate of 20.0%. Inset: the spectrum of pump laser. (c) Power dependence of OPO when pumping at 776.5 nm, showing a threshold pump power of ~22.5 μW and a gain rate of 18.7%. Inset: the spectrum of OPO. (d) The Fourier components of second order nonlinear coefficient

$d_{\text{eff}}$, which are ± 1942 (m$^{-1}$) and ± 5825 (m$^{-1}$). Inset: the variation of $d_{\text{eff}}$ of azimuth angle $\theta$.

**Fig. 1**

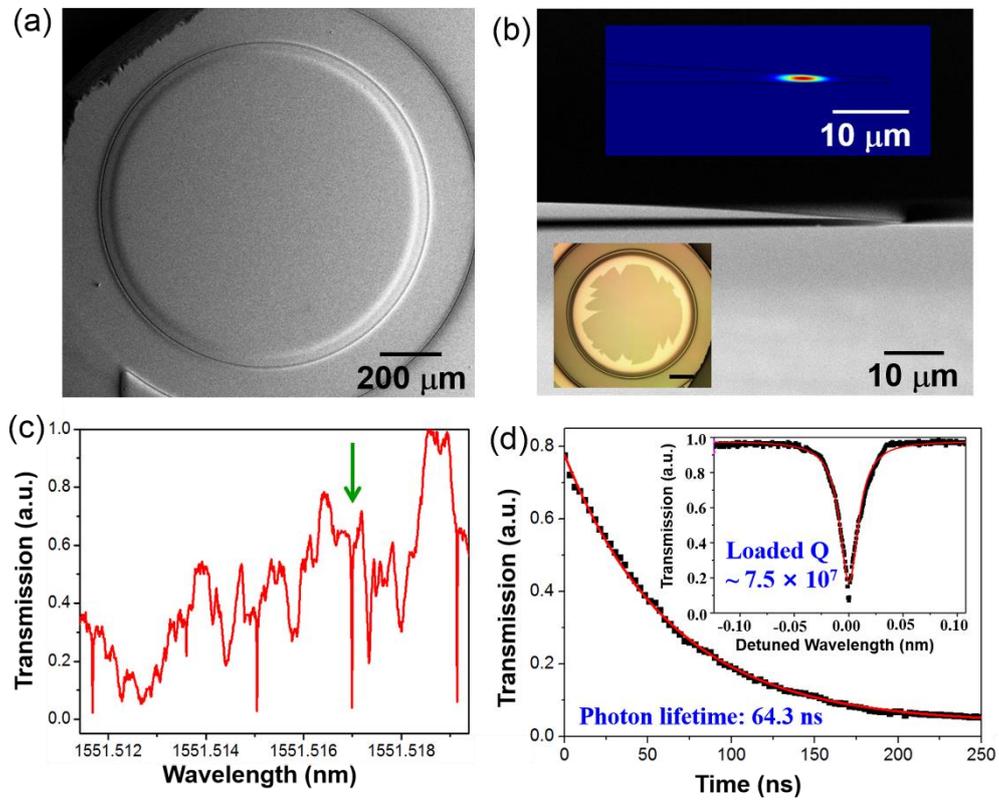

**Fig. 2**

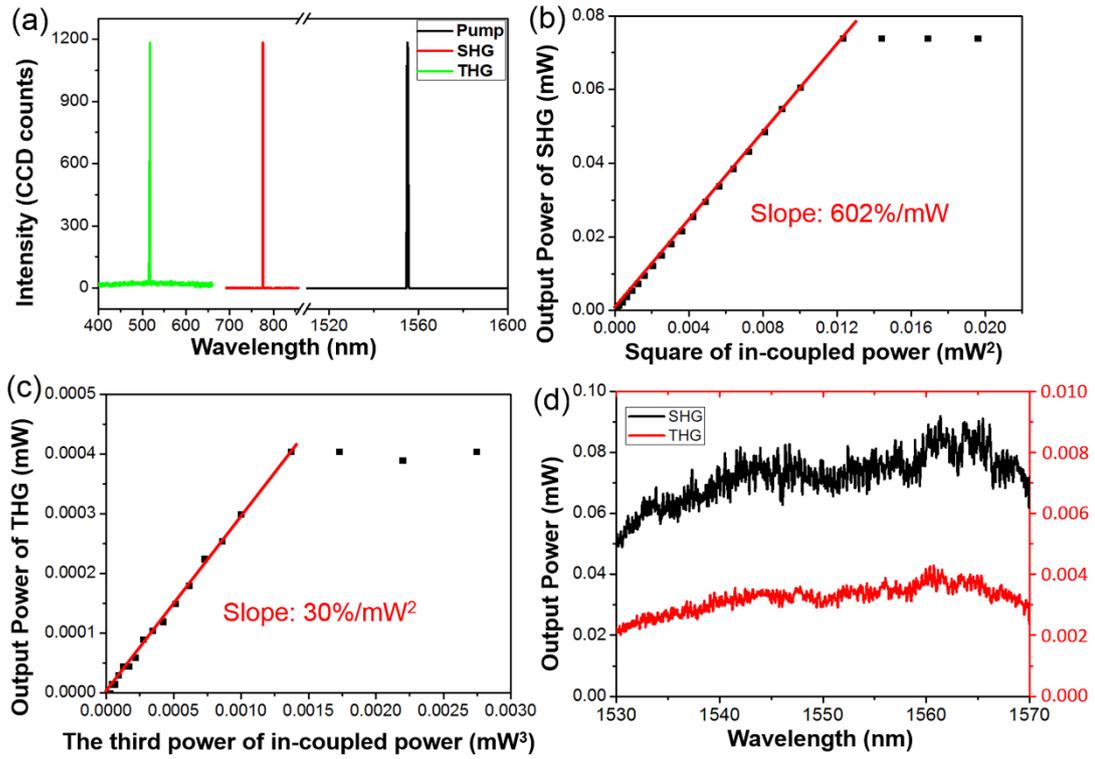

**Fig. 3**

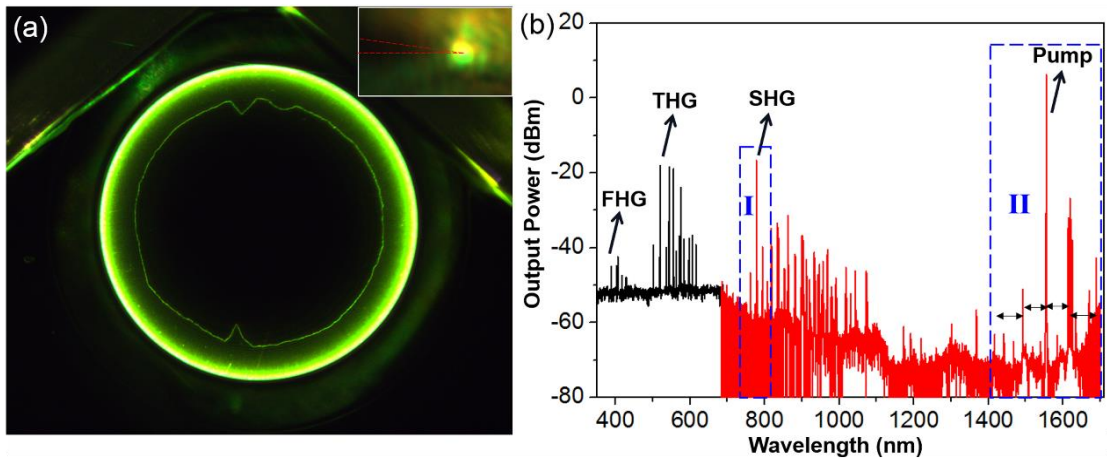

**Fig. 4**

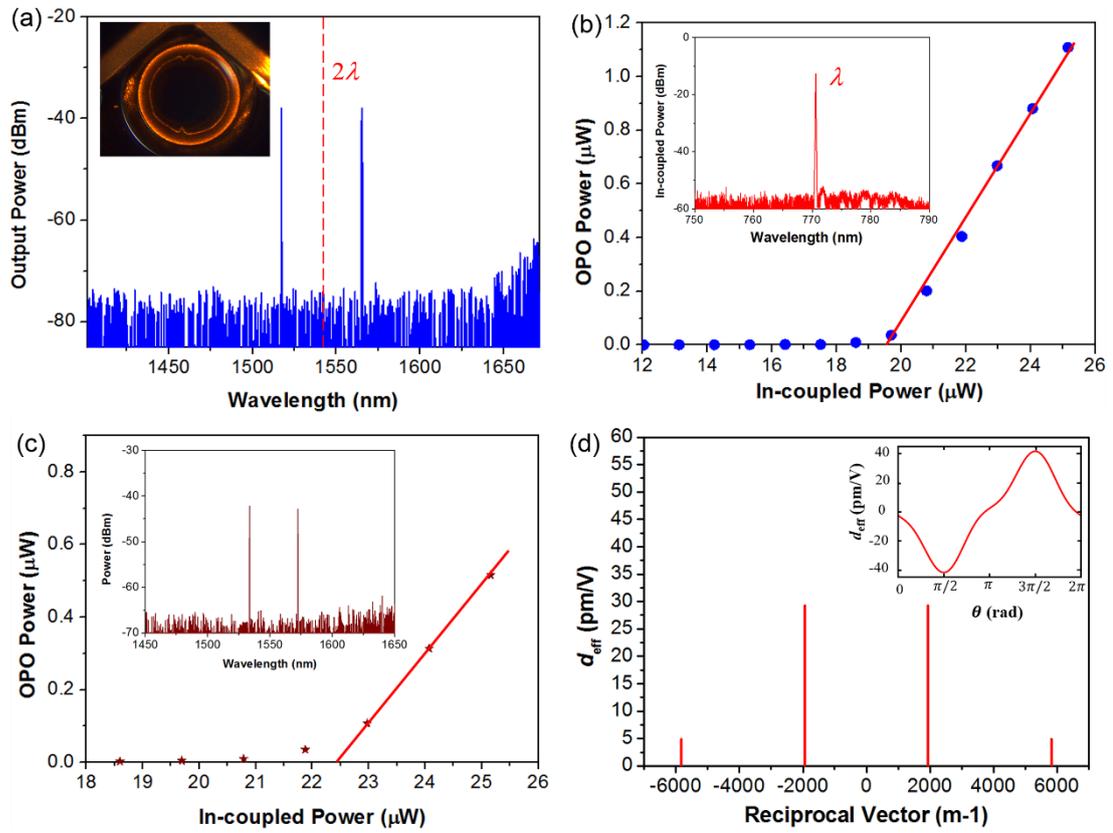

# Supplemental materials

In the Supplemental materials we provide the details of the fabrication process for ultrahigh-Q (UHQ) lithium niobate (LN) microdisk resonators, the experimental characterizations of LN microdisks with smaller diameters, as well as the discussion about the utility of natural quasi phase-matching (NQPM) for versatile and efficient nonlinear processes in UHQ micro-resonators.

## Fabrication process flow

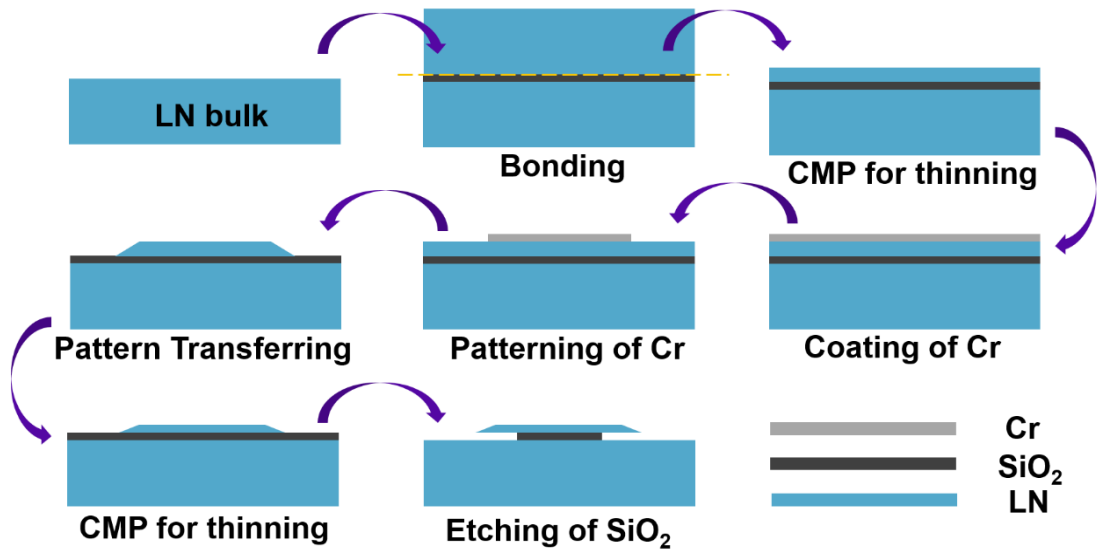

Fig. S1 Schematic illustration of fabrication process of UHQ microdisk on LN thin film.

The UHQ microdisk resonators were fabricated on a pristine LN thin film wafer with technological process schematically shown in Fig. S1. First, an X-cut LN thin film wafer without ion-slicing was prepared. Here, a monocrystalline LN bulk, which is bonded on anther LN substrate sandwiched with a silica transition layer, was thinned via chemo-mechanical polishing (CMP) to the thickness of more than 4 μm for the sake of maintaining wafer-scale homogeneity of surface planeness. Then the microdisk resonators were fabricated on such prepared LN thin film wafers by photolithography assisted by chemo-mechanical etching (PLACE) [1], which includes coating with chromium (Cr) layer, patterning of Cr layer with femtosecond laser lithography, etching of LN thin film by pattern transferring via CMP, a second thinning of the LN thin film microdisk via CMP, and final removal of the underneath silica layer into a small pillar by HF acid etching. Thereafter, a free-standing LN microdisk resonator with ~1.03 mm diameter was produced, as shown in Fig. 1 of the main text.

As mentioned in the main text, the essential advantage of the above fabrication process is on the optimized use of CMP instead of usually adopted ion-implantation and plasma etching for LN thin film preparation and etching. Recent works on LN photonic integrated components and circuits are exclusively using the thin film LN on insulator (LNOI) as the underlying material platform. LNOI is obtained by the ion-slicing of submicron-thickness films from the single-crystalline LN bulk, enabled by the implantation of helium ions ($He^+$) into the LN crystal to define a cleavage plane [2]. The crystalline quality of the ion-sliced LN film bonded on LN substrate had been previously examined using Rutherford backscattering (RBS), revealing an inferior quality of the ion-sliced thin film compared to the LN bulk crystal [3]. Moreover, the quality of the ion-sliced film can only be partially recovered to the virgin bulk crystal after high-temperature annealing above 1000 °C [3]. In practical photonic applications, the LN thin film is typically bonded on silica transition layer for high refractive index contrast. The annealing at this high temperature will destroy the bonding between the io-sliced film and the underneath silica layer [2]. Therefore, a relatively low temperature of 500 °C is chosen for annealing [4]. As a result, the inherent property of LNOI is inferior to the bulk LN crystal, rendering the attainable quality-factors of LNOI micro-resonators well below the resonators made from bulk LN crystals despite the utilization of high-precision patterning methods (like CMP) to shape the LNOI micro-resonators with surface/interface roughness comparable with the surface-tension induced atomic scale finish.

Geometrical designs also help in the UHQ performance of the fabricated LN microdisks. The thickness of the microdisk from the edge toward the center was varied from <1 μm to 3.5 μm. This transition was chosen to avoid collapsing of the microdisk, whose freestanding part was as wide as 100 μm, as shown in Figs. 1(a-b) of the main text. This is another key factor to protect the mode from the scattering loss by the undercut silica pillar. For comparison, a microdisk was fabricated on 700 nm thickness ion-sliced LN thin film (LNOI), as shown in Fig. S2. The freestanding part of this microdisk was chosen as narrow as 25 μm. However, due to the fragility of the submicron-thickness LN thin film, the microdisk was collapsed. Therefore, UHQ pristine LN thin film microdisks are successfully produced by avoiding the interior material defect (which is unavoidable during the ion-implantation in the ion-slicing process for LNOI preparation) and suppressing unfavorable interfaces via optimizing the fabrication process.

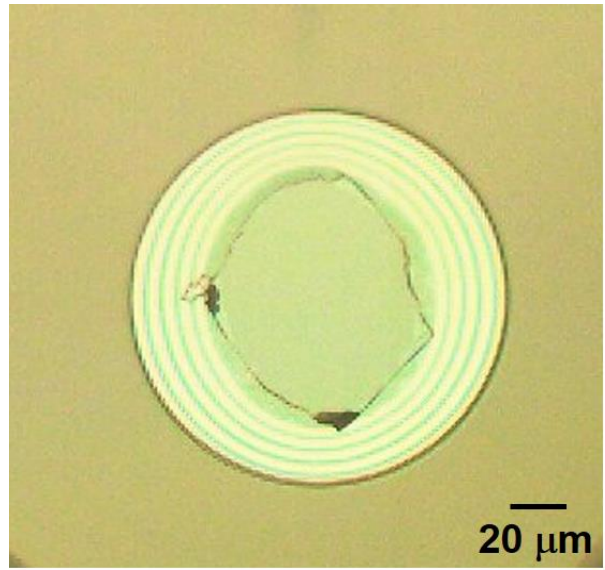

Fig. S2 Optical micrograph of a microdisk fabricated on ion-sliced thin film by PLACE.

## Characterizations of pristine LN microdisks with smaller diameters

Figure S3 shows the results for smaller microdisks fabricated on pristine LN thin film wafer. Though the radial widths of the freestanding part were more than 25 μm, the microdisks were intact. The microdisk with the diameter of 130 μm was shown in Fig. S3a, whose loaded Q factor was measured as $5.6\times10^7$ giving the intrinsic Q factor of $0.9\times10^8$. The other microdisk with the diameter of 240 μm is also shown in Fig. S3d, whose loaded Q factor was measured as $6.5\times10^7$ indicating intrinsic Q factor as $0.9\times10^8$. Since the bending losses are exponentially inverse proportional to diameters of the resonators, the tiny variance (<30%) of the Q-factors for microdisk resonators with diameters from ~1 mm to ~100 μm indicates the obtained Q-factors are solely determined by the intrinsic material absorption.

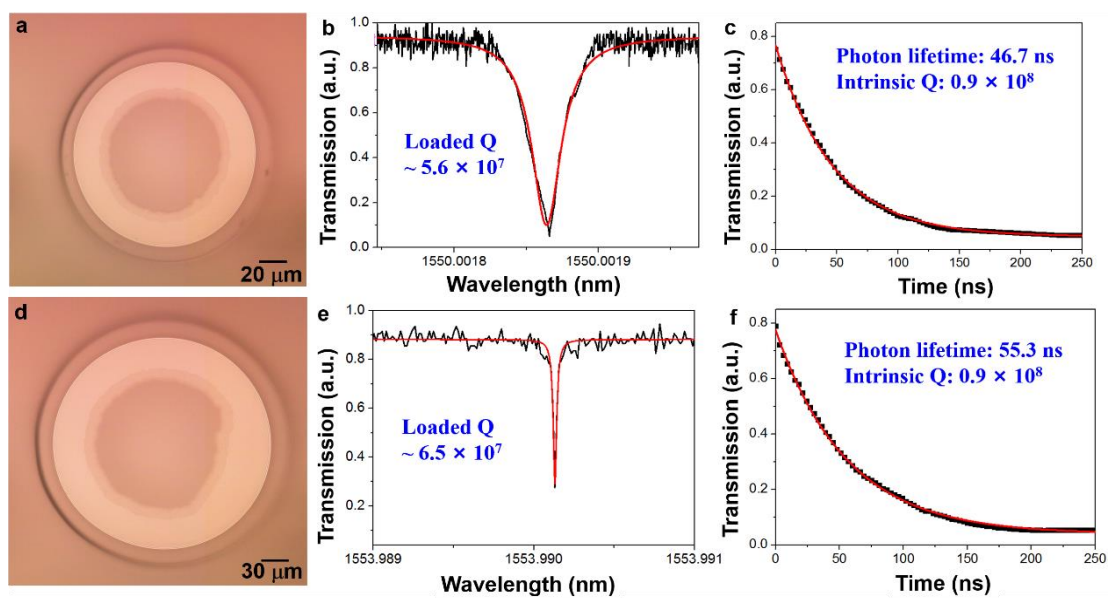

Fig. S3 The results for the smaller microdisks. (a) Optical micrograph of the microdisk with the diameter of 130 μm. (b)-(c), The loaded Q factor and the intrinsic Q factor of the microdisk shown in (a). (d) Optical micrograph of the other microdisk with the diameter of 240 μm. (e)-(f) The loaded Q factor and the intrinsic Q factor of the microdisk shown in (d).

# Utilization of natural quasi phase-matching for nonlinear optical processes

For an X-cut LN microdisk resonator, light loaded in the transverse-electric (TE) whispering gallery modes (WGM) will experience an alternating refractive index due to the birefringence of LN in the plane of the microdisk (see Fig. S4). The local refractive index of light circulating along the circumstance of the microdisk, is dependent on the azimuthal angle $\theta$ as:

$$\frac{1}{n^2(\lambda)} = \frac{\cos^2\theta}{n_o^2(\lambda)} + \frac{\sin^2\theta}{n_e^2(\lambda)} \quad (1)$$

where $n_o(\lambda)$ and $n_e(\lambda)$ are the effective ordinary and extraordinary refractive indices of modes at the wavelength of $\lambda$.

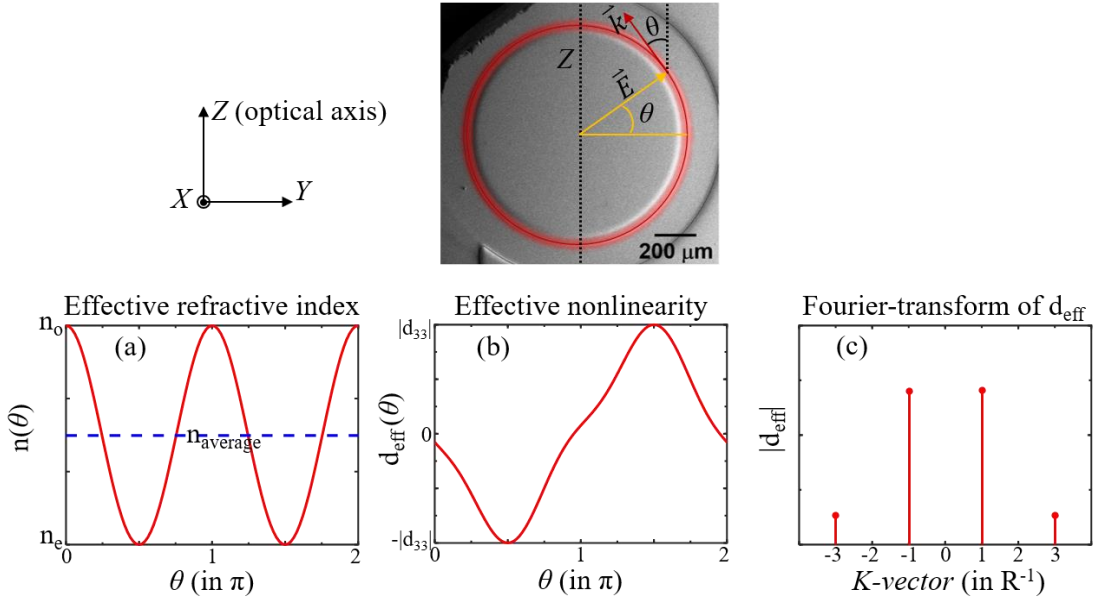

Fig. S4 The physical properties for X-cut LN microdisk. The LN microdisk is shown in the top panel, where the optical axis (z), the local wavevector ($\vec{k}$), the electric field vector ($\vec{E}$) and the azimuthal angle ($\theta$) are respectively labelled for the TE-modes. (a) The oscillation of effective refractive index (red solid line), and the refractive index averaged over one circle is plotted as the blue dashed line. (b) The alternation of effective nonlinearity. (c) The wavevector (*K*-vector) representation of $d_{\text{eff}}$, the horizontal unit is $R^{-1}$ ($R$ is the radius of the microdisk).

Similarly, the effective second-order nonlinear coefficient for the TE modes can be written as:

$$d_{eff}(\theta) = -d_{22} \cos^3 \theta + 3d_{31} \cos^2 \theta \sin \theta + d_{33} \sin^3 \theta \tag{2}$$

where $d_{ii}$ (*ii*=22, 31, 33) is the tensor components of the second-order nonlinear susceptibilities of bulk LN. Th angular variation of $d_{eff}(\theta)$ can be understood as an effective nonlinearity patterning (domain inversion) consisting of certain angular frequencies, which can be expressed in terms of a Fourier series as:

$$d_{eff}(\theta) = \sum_{m=-\infty}^{\infty} D_m e^{im\theta} \tag{3}$$

Combining equations (2) and (3) one can easily get the expression below:

$$d_{eff}(\theta) = D_1 e^{i\theta} + D_{-1} e^{-i\theta} + D_3 e^{3i\theta} + D_{-3} e^{-3i\theta} \tag{4}$$

where the coefficients $D_i$ (*i*=±1, ±3) are given by: $D_1 = -\frac{3}{8}(id_{31} + id_{33} + d_{22})$, $D_{-1} = \frac{3}{8}(id_{33} - d_{22} + id_{31})$, $D_3 = \frac{1}{8}(id_{33} - 3id_{31} - d_{22})$, $D_{-3} = -\frac{1}{8}(id_{33} - 3id_{31} + d_{22})$.

The angular variation of $d_{eff}(\theta)$ can be readily converted to the spatial variation $d_{eff}(s)$ (*s*=*Rθ* is the trajectory along the periphery of the microdisk, and *R* is the radius of the microdisk) as:

$$d_{eff}(s) = D_1 e^{\frac{i}{R}s} + D_{-1} e^{-\frac{i}{R}s} + D_3 e^{\frac{3i}{R}s} + D_{-3} e^{-\frac{3i}{R}s} \tag{5}$$

The spatial vectors of the nonlinearity patterning can be readily seen as $K_i = \pm \frac{1}{R}, \pm \frac{3}{R}$ (see also Fig. S4(c)). It is interesting to noted that the supported "poling" periods by natural quasi phase-matching are fixed at $\Lambda_i = \frac{2\pi}{|K_i|} = 2\pi R, 2\pi R/3$, correspond to the full and one thirds of the perimeter of the microdisk which are in turn determined by the symmetry of the microdisk as well as the LN crystal lattice.

The wave-vector mismatch $\Delta k$ for certain nonlinear process (i.e., second-harmonic generation, optical parametric oscillation, sum-frequency generation) is also alternating due to the variation of refractive index for the involved waves. For quasi phase-matching to be effective [5], one of the $K$-vectors by NQPM should be well within the oscillation range of $\Delta k$, indicating a flexible criterion for natural quasi phase-matching as:

$$\frac{1}{2\pi}\int_0^{2\pi} \Delta k(\theta)\, d\theta \equiv \Delta k_0 \approx K_i \equiv \pm\frac{1}{R}, \pm\frac{3}{R} \tag{6}$$

Fig. S5 demonstrates the physical meaning of the above criterion.

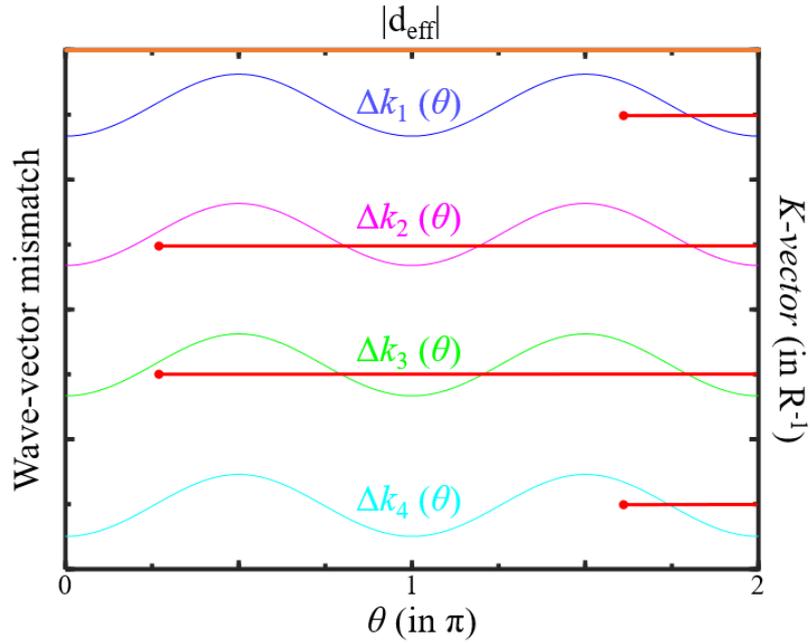

Fig. S5 The schematic for NQPM. The four oscillating curves of wave-vector mismatch $\Delta k(\theta)$ each encompasses one of the $K$-vectors by NQPM, and efficient frequency conversions are then expected.

**Reference:**

1. R. Wu, *et al.* Lithium niobate micro-disk resonators of quality factors above $10^7$. *Opt. Lett.* **43**, 4116-4119 (2018).

2. G. Poberaj, *et al.* Lithium niobate on insulator (LNOI) for micro-photonic devices. *Laser Photon. Rev.* **6**, 488-503 (2012).

3. P. Rabiei and P. Günter. Optical and electro-optical properties of submicrometer lithium niobate slab waveguides prepared by crystal ion slicing and wafer bonding. *Appl. Phys. Lett.* **85**, 4603–4605 (2004).

4. Y. Jia, et al. Ion-cut lithium niobate on insulator technology: Recent advances and perspectives. *Appl. Phys. Rev*. **8**, 011307 (2021).

5. J. Lin, et al. Broadband Quasi-Phase-Matched Harmonic Generation in an On-Chip Monocrystalline Lithium Niobate Microdisk Resonator. *Phys. Rev. Lett.* **122**, 173903 (2019).